\documentclass[man,floatsintext]{apa7}

\usepackage{lipsum}

\usepackage[american]{babel}
\usepackage{csquotes}
\usepackage[style=apa,sortcites=true,sorting=nyt,backend=biber]{biblatex}
\DeclareLanguageMapping{american}{american-apa}
\usepackage{graphicx}
\usepackage{booktabs,tabularx}
\usepackage{float}
\usepackage{amsmath}
\usepackage{amsfonts}
\usepackage{hyperref}

\title{The Universal Law of Generalization Holds for Naturalistic Stimuli}
\shorttitle{Universal Law Holds for Naturalistic Stimuli}

\author{Raja Marjieh\textsuperscript{1}, Nori Jacoby\textsuperscript{2}, Joshua C. Peterson\textsuperscript{3}, Thomas L. Griffiths\textsuperscript{1,3}}
\affiliation{\textsuperscript{1}Department of Psychology{,} Princeton University{,} USA, \textsuperscript{2}Max Planck Institute for Empirical Aesthetics{,} Germany, \textsuperscript{3}Department of Computer Science{,} Princeton University{,} USA}

\authornote{
  Corresponding author: Raja Marjieh. E-mail: \texttt{raja.marjieh@princeton.edu}}
\abstract{Shepard's universal law of generalization is a remarkable hypothesis about how intelligent organisms should perceive similarity. In its broadest form, the universal law states that the level of perceived similarity between a pair of stimuli should decay as a concave function of their distance when embedded in an appropriate psychological space. While extensively studied, evidence in support of the universal law has relied on low-dimensional stimuli and small stimulus sets that are very different from their real-world counterparts. This is largely because pairwise comparisons -- as required for similarity judgments -- scale quadratically in the number of stimuli. We provide direct evidence for the universal law in a naturalistic high-dimensional regime by analyzing an existing dataset of $214,200$ human similarity judgments and a newly collected dataset of $390,819$ human generalization judgments ($N=2406$ US participants) across three sets of natural images.}


\begin{document}
\maketitle

\section{Statement of Relevance}
Humans constantly form generalizations, whether when trying to identify the color of an object or reasoning about which action to take based on past experiences. Understanding how generalizations relate to underlying psychological representations is a core problem in cognitive science. The universal law of generalization is a fundamental hypothesis concerning the nature of this relationship which states that the strength of generalization between two stimuli should decay as a universal exponential function of their psychological distance. While extensively studied, evidence for the universal law comes from small datasets and artificial stimuli that are very different from the real world. Our work is the first to provide direct evidence for the universal law in a high-dimensional naturalistic domain by collecting and analyzing $605,019$ human similarity and generalization judgments for natural images.  
%390,819 + 71,400 * 3
\newpage

%\section{Introduction} %TLG no title for introduction section

Every day, humans interact with complex perceptual objects that vary in modality, structure and function. Whether deciding when to cross a street, recognizing the face of a friend, or trying to determine whether a novel fruit will taste good, we need to form meaningful generalizations from past perceptual experiences. This problem of generalization is arguably one that we share with all intelligent species, something that led Roger Shepard to propose a candidate for the first universal law of psychology \parencite{shepard1987toward}. Shepard's universal law of generalization -- intended to hold for intelligent entities anywhere in the universe -- asserts that the extent to which a property is generalized from one stimulus to another should decrease as a concave function (usually exponential) of the distance between those stimuli in psychological space. This idea has been elaborated upon in Bayesian models of cognition \parencite{tenenbaum_griffiths_2001}, and linked to information-theoretic principles such as maximum entropy \parencite{myung1996maximum}, Kolmogorov complexity \parencite{chater2003generalized}, and efficient coding \parencite{Sims2018}.

Implicit in Shepard's proposal is the idea that it is possible to represent perceptual stimuli in a psychological space -- typically a low-dimensional representation where the similarity between two stimuli decreases with their distance. While this idea is controversial (e.g., \cite{tversky1977features}; \cite{tversky1986nearest}; \cite{peer2021structuring}), Shepard showed that such spaces can capture the similarity relationships between a variety of simple perceptual stimuli. He proposed a procedure, known as multidimensional scaling (MDS), for uncovering the structure of mental representations from behavioral data (\cite{shepard1962analysis}; \cite{shepard1980multidimensional}; \cite{steyvers2002multidimensional}). Given a set of stimuli, the procedure begins by constructing a similarity matrix between all stimulus pairs, e.g., by collecting similarity judgments or confusion probabilities, and then applying an iterative algorithm that embeds those stimuli in a psychological space (typically Euclidean) such that similar stimuli are mapped to nearby points. 

Having mapped stimuli to points in a psychological space, it becomes possible to test the universal law. While Shepard's (non-metric) multidimensional scaling method assumes similarity decreases with distance, it doesn't specify the form of that function. By analyzing the abstract question of how an ideal organism should decide whether two stimuli shared a given property, \textcite{shepard1987toward} showed that this function should be concave. Mathematical analysis of a variety of different assumptions about the distribution of properties in psychological space showed that generalization typically decreased as an exponential function of distance. Shepard then demonstrated that this theoretical relationship held for a wide array of stimuli that had been embedded into a psychological space via MDS, including geometric shapes, phonemes, colors (in both humans and pigeons), and even Morse code signals. 

%Using this procedure, Shepard revealed a surprising property of mental representations by providing compelling evidence for a universal law that links similarity gradients and distances in psychological space  \parencite{shepard1987toward}. In its most general form, the universal law of generalization states that the level of similarity between stimuli should decay as a concave function (usually exponential) of their distance in psychological space 

Despite the success of Shepard’s account, two clear limitations remain. First, for a set of $N$ stimuli MDS requires on the order of $N^2$ pairwise comparisons to construct a full similarity matrix which, as the number of stimuli increases, necessitates a large amount of human data. For example, a set of 100 stimuli would require on the order of 10,000 similarity judgments, without even including any repetitions to ensure data quality. This bottleneck has recently propelled a line of research aimed at finding cheaper approximations for human similarity matrices (\cite{roads2021enriching}; \cite{jha2023extracting}; \cite{marjieh2022words}). Second, and in part as a result of the first limitation, most of the evidence for the universal law comes from studies that are limited to low-dimensional artificial stimuli and small stimulus sets (\cite{shepard1987toward}; \cite{cheng2000shepard}; \cite{Ghirlanda200315}; \cite{Sims2018}). %Indeed, the stimuli considered in \textcite{shepard1987toward} were basic geometric shapes such as circles and triangles, monochromatic colors, simple phonemes and Morse code signals. Likewise, 
Even though more recent work such that of \textcite{Sims2018} has considered somewhat richer stimuli such as synthesized instrument timbres and vibrotactile patterns, these were still limited to small datasets on the scale of 10-20 stimuli. These limitations make it hard to draw conclusions about the status of the universal law of generalization in the high-dimensional regime of real-world stimuli, especially as fundamental problems in psychology continue to be reshaped by large-scale behavioral studies (see e.g., \cite{awad2018moral}; \cite{battleday2020capturing}; \cite{peterson2021using}; \cite{marjieh2022reshaping}). 

To address this gap, we leveraged recent advances in online recruitment as well as the availability of naturalistic image datasets to directly test the universal law of generalization in a high-dimensional setting. Specifically, we considered a dataset of similarity judgments over three sets of images recently collected by \textcite{Peterson2018} where each dataset comprised 120 images from a given natural category, namely, animals, fruits, and vegetables. This dataset consisted of 214,200 human judgments. To account for the different ways in which similarity scores can be constructed, we augmented this dataset with a newly collected set of generalization judgments where participants rated how likely it is a certain \textit{blank property} (\cite{osherson1990category}; \cite{kemp2009structured}) (e.g., having an enzyme) generalizes from one stimulus to another. The latter dataset comprises  $390,819$ generalization judgments from $2,406$ online participants. We used these data to directly test the universal law of generalization in this high-dimensional large-scale regime.

All data and analysis code considered in the present work, as well as all necessary code for reproducing the online behavioral experiments are made publicly available in the following OSF repository: \url{https://tinyurl.com/tmwu6pxv}. All participants provided informed consent prior to participation in accordance with an approved Princeton University Institutional Review Board (IRB) protocol (\#10859). 

% \subsection{Previous Work}
% The original paper in which the universal law was first postulated already contained a wide array of evidence in its support. To buttress his claim, Shepard analyzed human and pigeon data from studies that involved a variety of stimuli, including simple shapes such as circles and triangles, colors, phonemes, and Morse code signals. Specifically, given a matrix of generalization probabilities, $p_{ij}$, where each entry corresponds to the probability of generalizing from some stimulus $i$ to another $j$, Shepard computed a generalization score matrix $g_{ij}=[(p_{ij}p_{ji})/(p_{ii}p_{jj})]^{1/2}$ and then applied the non-metric MDS algorithm to construct an embedding of the stimuli in psychological space. Then, by computing generalization gradients, i.e., by plotting the generalization scores against distances in psychological space, Shepard showed that all data points fell nicely on a universal exponential curve. More research followed since the original study (Ghirlanda and Enquist, 2003; Sims, 2018), extending the validity of the law to somewhat richer stimuli, such as synthesized instrument timbres and vibrotactile patterns, as well as to other species (e.g. honeybees, Cheng, 2000). To the best of our knowledge, however, no study to this point has provided direct evidence for the law over fully-fledged real-life stimuli such as natural images. 

\section{Methods}

\begin{figure}[htp]
  \centering
  \includegraphics[width=\linewidth]{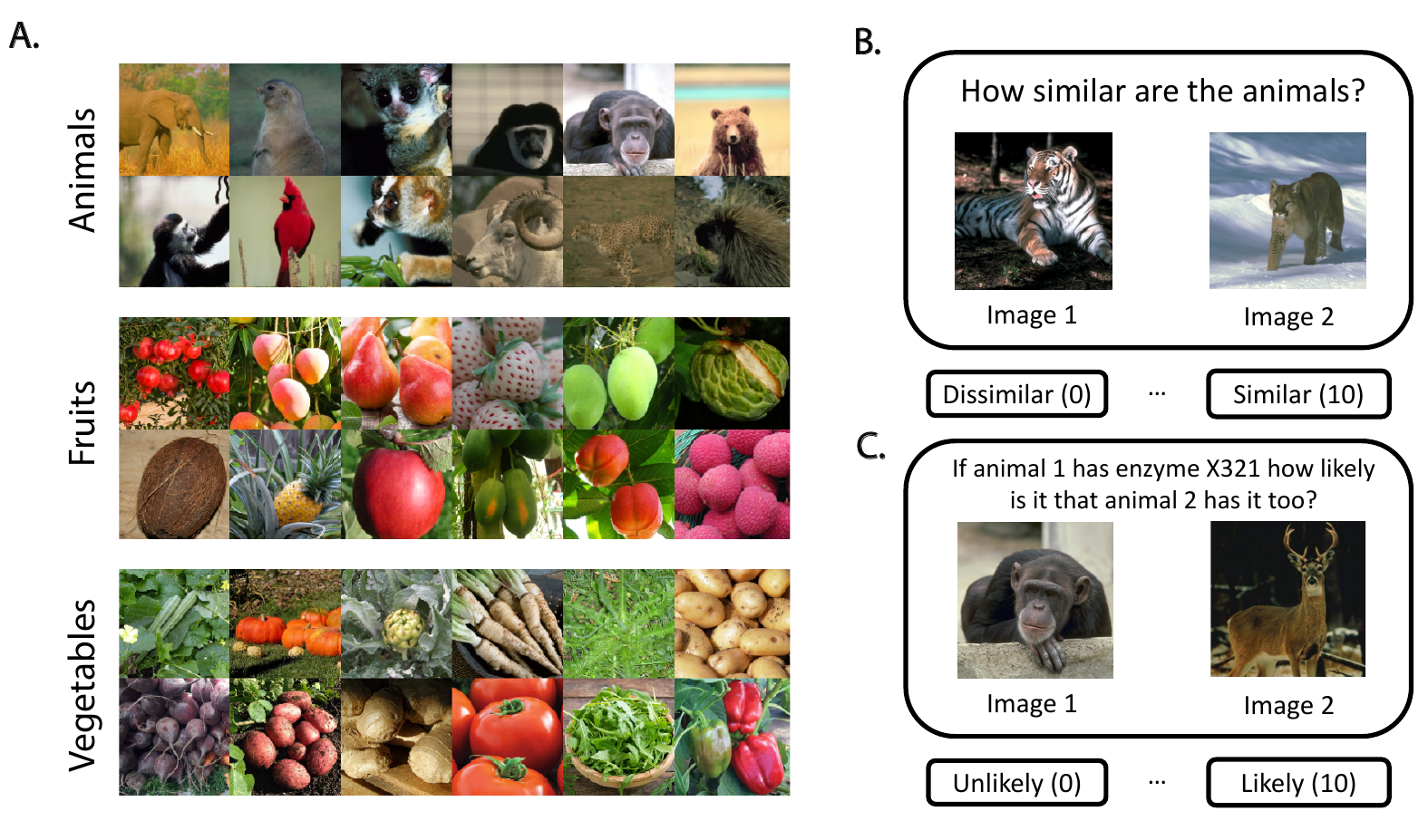}
%   \vspace{-4mm}
  \caption{Example images from three natural categories, namely, animals, fruits, and vegetables (\textbf{A}), and schematics of the two elicitation queries used in the present work, namely, direct similarity judgments (\textbf{B}), and generalization judgments (\textbf{C}).}
  \label{fig:schematic}
\end{figure}
Our approach builds on advances in large-scale online recruitment and experiment design to exhaustively estimate similarity matrices over naturalistic stimuli by directly scaling up pairwise judgment elicitation. For stimuli, we focused on natural images (Figure~\ref{fig:schematic}A) for three reasons, namely, a) they strike a balance between being perceptually complex and being intuitive and widespread across cultures, b) they can be easily embedded within an online study, which facilitates crowdsourcing, and c)  high-quality sets of natural images along with accompanying behavioral data are available in the literature (\cite{peterson2017evidence}; \cite{Peterson2018}; \cite{jha2023extracting}). As for the paradigm, we used simple pairwise judgment elicitation on a Likert scale with two complementary types of human judgments that are common to the study of representations, namely, direct similarity judgments that answer the query ``How similar are the animals in the following two images?'' (Figure~\ref{fig:schematic}B; \cite{shepard1980multidimensional}; \cite{Peterson2018}), and generalization judgments that answer the query ``If the animal in image 1 has enzyme X321, how likely is it that the animal in image 2 has it too?'' (Figure~\ref{fig:schematic}C; \cite{osherson1990category}; \cite{kemp2009structured}). In the latter case, we used a fictitious enzyme name so as to prevent participants from resorting to any technical knowledge.
Recent work on similarity judgments for images has also used an alternative paradigm where participants judge which of three images is the odd one out \parencite{hebart2020revealing}. However, we chose to use pairwise similarity judgments to maximize the correspondence with the paradigm used by Shepard, and because triplets scale cubically in the number of stimuli making them even harder to scale (without further assumptions about deriving pairwise similarity from triplets). 

\subsection{Stimuli}
We used sets of images from three natural categories, following \textcite{Peterson2018}. Each set comprised 120 images from one of the following categories: animals, fruits, and vegetables (see examples in Figure~\ref{fig:schematic}A). In addition, these datasets were supplemented with full $120\times120$ symmetric human similarity matrices $s_{ij}$ where each entry corresponds to an aggregate similarity score between an image $i$ and an image $j$ in the range $0-1$, where a value of $0$ indicates complete dissimilarity, and a value of $1$ indicate complete similarity. Each such similarity matrix was constructed using 71,400 human judgments from a pool of approximately 1,200 US participants recruited on Amazon Mechanical Turk (AMT) \parencite{Peterson2018}.

\subsection{Participants}
Participants for the generalization tasks were recruited online via AMT subject to the following criteria to ensure data quality: 1) participants must be at least 18 years of age, 2) they must reside in the United States, and 3) they must have an approval rate of 95\% or higher on AMT. The recruitment process was performed using the Dallinger\footnote{\url{https://dallinger.readthedocs.io/en/latest/}} platform for experiment hosting, and the experimental session was programmed using PsyNet, a framework for online experiment design that is built on top of Dallinger \parencite{HarrisonMarjieh2020}. Overall $N = 2406$ participants completed the studies, and they were paid \$12/hour for their participation. Specifically, the $N=773$ participants in the animals condition had an age range of $21-78$ years ($M=38.3$, $SD=11.0$), the $N=833$ participants in the fruits condition had an age range of $19-70$ years ($M=39.3$, $SD=11.0$), and the $N=800$ participants in the vegetables condition had an age range of $20-77$ years ($M=39.1$, $SD=10.9$). The sample size was selected such that each image pair received an average of 9 ratings to match that of the similarity datasets of \textcite{Peterson2018}.

% mention just age range.
\subsection{Procedure}
After completing a consent form, participants received the following instructions “In this experiment we are studying how people form generalizations. In each trial of this experiment you will be presented with two images of animals / fruits / vegetables. One of the animals / fruits / vegetables will possess a certain property, and your task will be to judge based on that information how likely it is that the second animal / fruit / vegetable has that property. You will have eleven response options, ranging from 0 (‘Not Likely at All’) to 10 (‘Very Likely’). Choose the one you think is most appropriate”. Participants then proceeded to the main experiment where they were presented with image pairs followed by the prompt “If the animal / fruit / vegetable in the left image has enzyme X132, how likely is it that the animal / fruit / vegetable in the right image has it too?” (see schematics in Figure~\ref{fig:schematic}C). Overall, 390,819 judgments were elicited with each each participant providing up to 200 judgments. The procedure in the similarity paradigm of \textcite{Peterson2018} was analogous. Participants rated the similarity between pairs of images on a Likert scale ranging from 0 (‘Not Similar at All’) to 10 (‘Very Similar’) (Figure~\ref{fig:schematic}B; see \cite{Peterson2018} for additional details). 

\subsection{Data Analysis}
\noindent\textbf{From generalization to similarity.}
To convert generalization scores into similarity matrices the following preprocessing was applied. First, the responses of individual participants were z-scored (within participant) to account for different usage of the response scale across participants. Then, the z-scored ratings were averaged across participants to produce a single score per stimulus pair. The summarized z-scores were then converted into generalization probabilities $p_{ij}$ by passing them through a cumulative normal distribution. Finally, to derive symmetric similarity matrices $s_{ij}$ we applied Shepard's similarity formula $s_{ij} = \sqrt{p_{ij}p_{ji} / p_{ii}p_{jj}}$ \parencite{shepard1987toward}.\footnote{Note that this formula does not change the similarity matrices of \textcite{Peterson2018} since $\sqrt{s_{ij}s_{ji} / s_{ii}s_{jj}}=\sqrt{s_{ij}^2}=s_{ij}$ due to the fact that $s_{ij}=s_{ji}\geq0$ and that $s_{ii}=1$.} In practice, we noticed that a few of the diagonal probabilities $p_{ii}$ were smaller than their off-diagonal counterparts which resulted in a generalization score that is greater than 1 and hence a negative entry in the distance (dissimilarity) matrix ($\Delta_{ij}=1-s_{ij}$), likely due to noise in the similarity estimates. Since these entries constitute only extremely small fraction of the data (0.8\%), we truncated the diagonal values by setting $p_{ii}$ to one, similar to \textcite{Peterson2018}.

\noindent\textbf{Multidimensional scaling.} Given a dissimilarity matrix $\Delta_{ij}$, MDS embeddings $z$ were obtained using the \verb|manifold.MDS| method from the scikit-learn Python library \parencite{scikit-learn} with a maximum iteration limit of 10,000 and a convergence tolerance of 1e-100. Embeddings were computed in two steps: first metric MDS was applied to get an initial embedding which was then used to initialize non-metric MDS. We chose a $d=4$ dimensionality for the embedding space based on an MDS stress curve analysis (shown in Supplementary Figure \ref{fig:stress-curves}; for visualization purposes only we used $d=2$ in the figures below) whereby the first dimension $d$ for which all stress values across all datasets dropped below $0.2$ was selected (a standard threshold above which MDS fit is deemed poor; \cite{kruskal1964multidimensional}). Finally, to construct generalization gradients we computed Euclidean distance between all MDS embedding vectors $d_{ij}=||z_i-z_j||_2$ and combined them with their corresponding similarity scores $s_{ij}$ to produce the two-dimensional set $\mathcal{D}=\{(d_{ij},s_{ij})\}$. We analyzed the resulting generalization gradients in two complementary ways, namely, by directly fitting curve models to the raw set $\mathcal{D}$, and by fitting them to a binned version of $\mathcal{D}$. The former served as a conservative test, and the latter as more balanced one meant to evaluate the average curve and to take into account the fact that different regions of the generalization gradient have different densities (e.g., high similarity pairs are much less common than low similarity pairs which can over-emphasize the tail of the gradient). The binning was done by dividing the distances $\{d_{ij}\}$ into 100 bins and then computing the average $d_{ij}$ and $s_{ij}$ within each bin and their standard errors. 

\begin{figure}[htp]
  \centering
  \includegraphics[width=\linewidth]{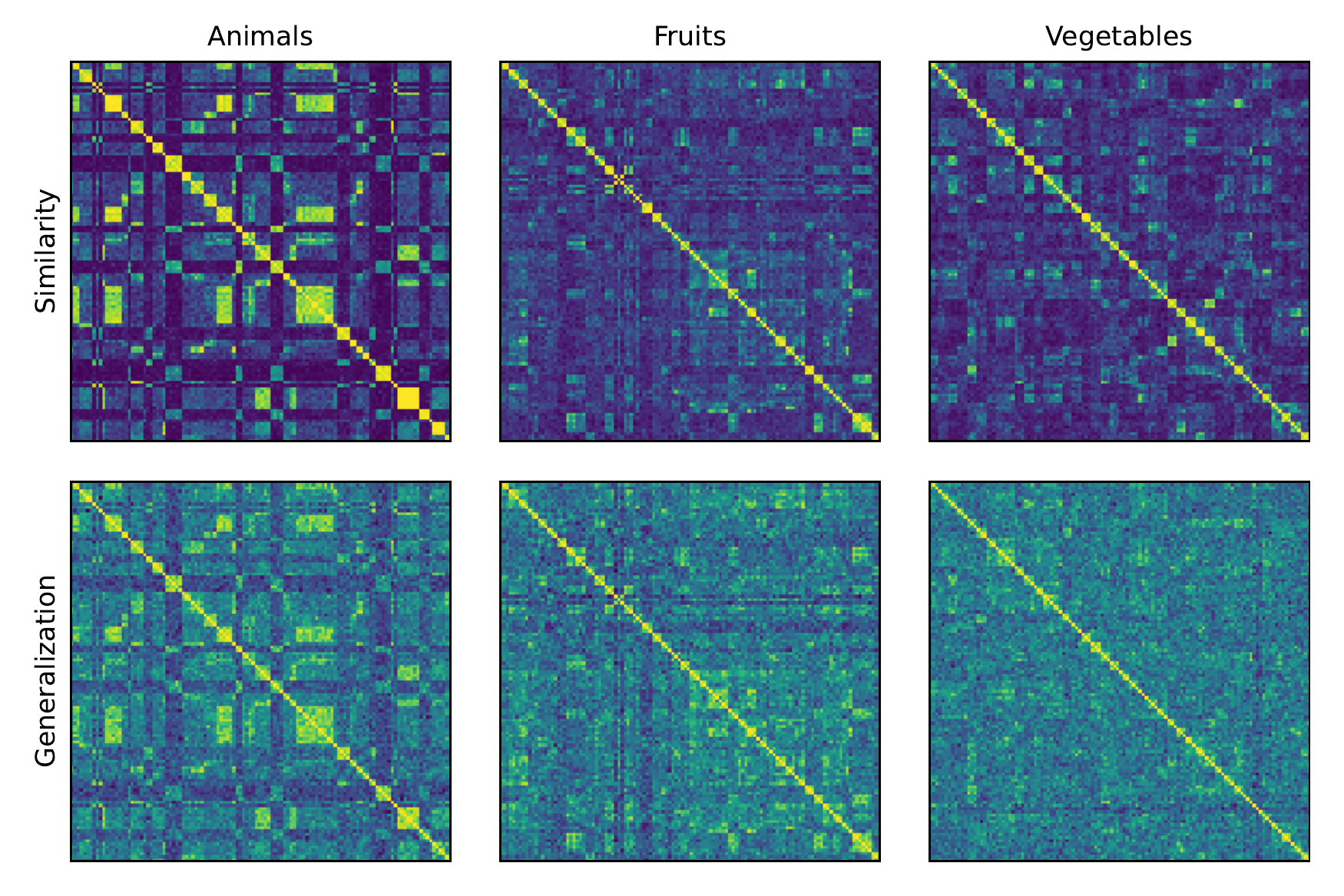}
%   \vspace{-4mm}
  \caption{Similarity matrices over the different domains of natural images and the two judgment elicitation tasks considered, namely, direct similarity judgments (top) and generalization judgments (bottom).}
  \label{fig:sim-mats}
\end{figure}

\noindent\textbf{Model fitting and evaluation.} To test the universal law, we evaluated the extent to which an exponential function of the form $g(x)=a e^{-b x} + c$ could account for the generalization gradients $\mathcal{D}$ relative to three other models of increasing complexity, namely, a simple linear model $g(x)=a x + b$, a quadratic model $g(x) = a x^2 + b x + c$ (same complexity as exponential but with the option of being either concave or convex), and a flexible generalized additive model (GAM), i.e., a model of the form $g(x)=\alpha+\sum_i\beta_if_i(x)$ where $f_i(x)$ is a basis of cubic splines \parencite{hastie2009elements}, as well as the intrinsic inter-rater variability of the data. To fit the exponential, linear, and quadratic models we used the \verb|curve_fit| least squares optimizer in \verb|scipy|, and to fit the GAM we used the \verb|LinearGAM| method of the \verb|pygam| package \parencite{serven2018pygam} which by default uses 20 basis functions and optimizes the model parameters in a grid-search manner. For model evaluation, and to accommodate both for the possibility of over-fitting and to adjust for degrees of freedom, we performed a split-half bootstrap analysis whereby 100 data splits were produced by randomly dividing the ratings per image pair in half and then producing two generalization gradients $\mathcal{D}_{h_1}$ and $\mathcal{D}_{h_2}$ to which the model was fitted yielding two sets of predictions $\{s^\prime_{ij}\}_{h_1}$ and $\{s^\prime_{ij}\}_{h_2}$. We then computed the following Pearson correlation coefficients between the data-model sets $\{s_{ij}\}_{h_1}$, $\{s^\prime_{ij}\}_{h_1}$ and $\{s_{ij}\}_{h_2}$, $\{s^\prime_{ij}\}_{h_2}$: $r_{dd}$ data-data correlation, $r_{mm}$ model-model correlation, $r_{dm}$ data-model correlation (there are two splits for each randomized split half, and thus there are two ways to compute this which we averaged), and $r_c=r_{dm} / \sqrt{r_{dd}r_{mm}}$ the data-model correlation corrected for attenuation. In addition, we provide the coefficient of determination (variance explained) $R^2$ for each of the fitted models, as well as, the Bayesian Information Criterion (BIC) for model selection which trades-off model complexity with goodness of fit \parencite{priestley1981spectral}. We used the Gaussian BIC formula which is given by $\text{BIC}=n\log(\text{RSS}/n)+k\log n$ where $\text{RSS}=\sum_i(x_i-\hat{x}_i)^2$ is the residual sum of squares between data and model, $n$ is the number of data points and $k$ is the number of fitted parameters. Finally, since only the relative BIC score is meaningful and because we are specifically interested in how models perform relative to the exponential solution, we report BIC scores relative to that solution $\Delta \text{BIC}=\text{BIC}_{\text{model}}-\text{BIC}_{\text{exponential}}$. We note that in computing all metrics we excluded trivial self-similarity points $(d=0,s=1)$ to prevent artificial inflation of values.

\section{Results}
\begin{figure}[htp]
  \centering
  \includegraphics[width=\linewidth]{figures/mds-2d-combined_v5.pdf}
  \vspace{-8mm}
    \caption{Two-dimensional MDS embeddings for the similarity and generalization data with the raw image stimuli from the different natural categories overlaid.}
  \label{fig:mds-2d}
\end{figure}

The average similarity matrices over the different domains and tasks are summarized in Figure~\ref{fig:sim-mats}. The first thing to observe is that the generalization and similarity judgment tasks yield results that are significantly correlated across domains, with a Pearson correlation of $r=.71$ (95\% CI $[.69,.74]$) for animals, $r=.55$ (95\% CI $[.52,.58]$) for fruits, and $r=.36$ (95\% CI $[.33,.40]$) for vegetables (CIs bootstrapped over participants with 100 repetitions). This is consistent with the expectation that generalization over blank properties and similarity judgments capture shared variance \parencite{kemp2009structured}. 

% \begin{figure}[htp]
%   \centering
%   \includegraphics[width=\linewidth]{figures/universal_law_multiple_models.pdf}
% %   \vspace{-4mm}
%   \caption{Raw generalization gradients across domains of natural images and tasks with the optimal exponential model plotted in dashed red lines.}
%   \label{fig:universal-law}
% \end{figure}

\begin{figure}[htp]
  \centering
  \includegraphics[width=\linewidth]{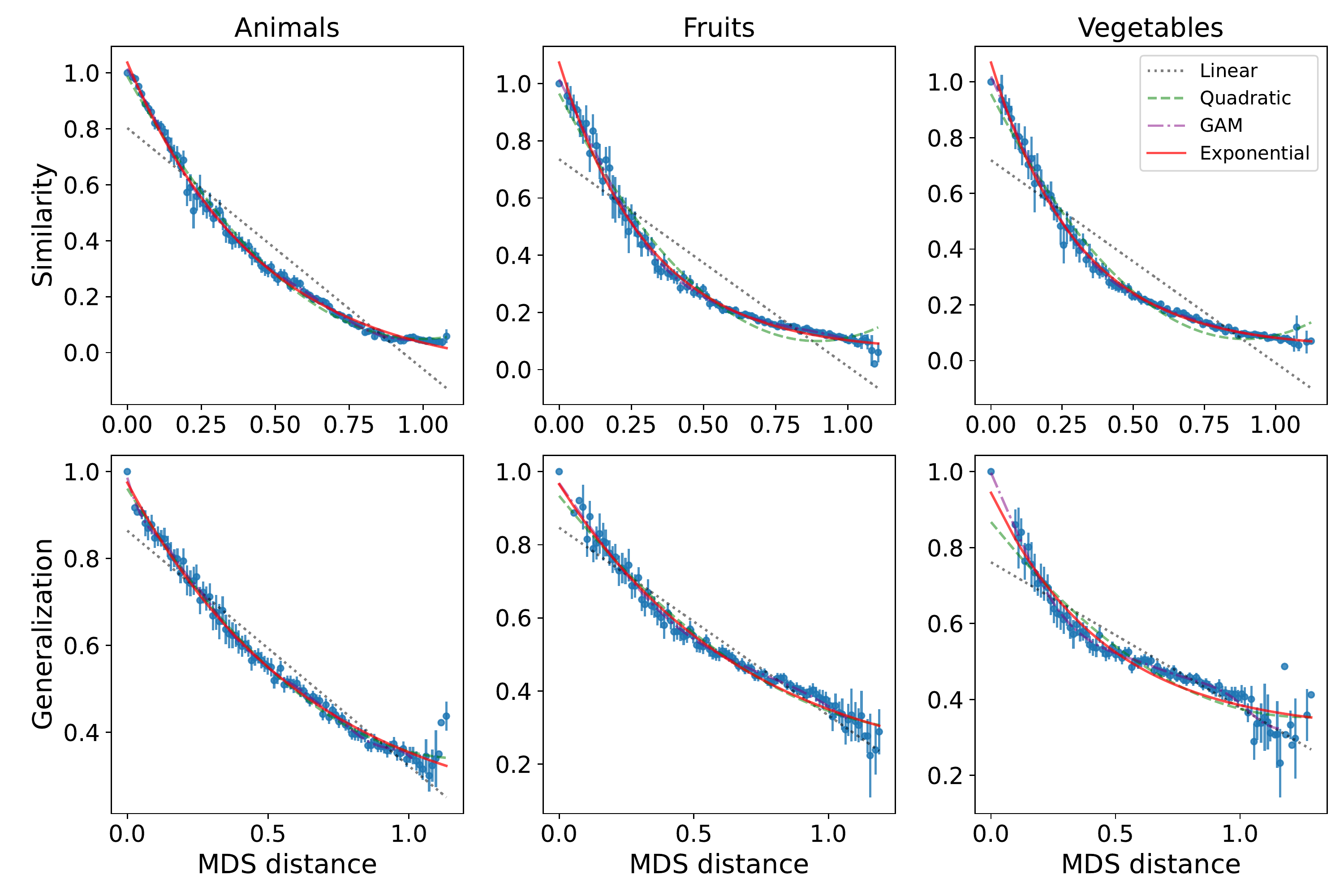}
%   \vspace{-4mm}
  \caption{Generalization gradients across domains of natural images and tasks with the optimal model fits overlaid. Error bars indicate 95\% confidence intervals.}
  \label{fig:universal-law}
\end{figure}

Next, to get a better sense of the psychological content of those spaces, we visualized their two-dimensional MDS solutions in Figure~\ref{fig:mds-2d}. All three domains revealed a semantically structured organization of the stimuli. In the case of animal images, distinct and interpretable organization schemes emerged, corresponding to animal categories such as herbivores, carnivores, amphibians, reptiles, and birds. In the case of fruits and vegetables the distribution is more continuous, with color serving as a clear semantic axis, with interpretable subclasses occupying different areas of the space such as citrus fruits and berries in the case of fruits, and whether the vegetable grows above or below the ground in the case of vegetables. These results are consistent with the findings of \textcite{Peterson2018} in the case of similarity and extend them to generalization, implying again that both tasks are capturing shared semantic content.

We are now ready to analyze the generalization gradients for each of the conditions and test to what extent they can be explained by the different models. The average binned gradients along with their optimal fit for all models are shown in Figure~\ref{fig:universal-law} (see Methods; raw gradients are shown in Supplementary Figure~\ref{fig:universal-law-raw}; explicit fitted parameter values and their CIs are provided in Supplementary Tables~\ref{tab:model-parmeters}-\ref{tab:model-parmeters-quadratic} and \ref{tab:model-parmeters-binned}-\ref{tab:model-parmeters-quadratic-binned} for the raw and binned analyses, respectively). As can be seen, the scatter points appear to follow a concave trend, with the similarity data in particular tightly tracking the exponential curve, which also overlaps substantially with the quadratic and GAM models. The linear model, on the other hand, appears to find some intermediate compromise due to its limited flexibility. To quantify this, we provide the full list of evaluation metrics on the raw gradients in Tables~\ref{tab:model-performance-similarity}-\ref{tab:model-performance-generalization} (see Supplementary Tables~\ref{tab:model-splithalf-similarity}-\ref{tab:model-splithalf-generalization} for additional metrics; see also Tables~\ref{tab:model-performance-similarity-binned}-\ref{tab:model-performance-generalization-binned} and \ref{tab:model-splithalf-similarity-binned}-\ref{tab:model-splithalf-generalization-binned} for the binned equivalent). The exponential function provides an excellent model for the data with an average model-data Pearson correlation of $r_c=.96$ (corrected for attenuation, see Methods). To see where the exponential model stands with respect to the different models in each condition, we bootstrapped the difference in variance explained $\Delta R^2 = R^2_\text{exponential}-R^2_\text{model}$, and the Bayesian Information Criterion $\Delta \text{BIC}$ (see Methods) between the exponential model and the other models. \\
Starting from the domain of similarity, we found that the exponential model outperformed the linear model in all three domains, with $\Delta R^2$ 95\% CIs given by $[.04,.10]$, $[.13,.14]$, $[.13,.14]$ for animals, fruits, and vegetables, respectively. The same pattern holds for the corresponding $\Delta\text{BIC}$ CIs (i.e., when penalizing for complexity), $[2644, 2921]$, $[1830, 2110]$, $[2185,2458]$ (positive values favor exponential in our definition of $\Delta \text{BIC}$). As for the quadratic solution, the models performed practically the same (with a slight boost for the exponential) with $\Delta R^2$ CIs given by $[-.001,.001]$, $[.009,.017]$, and $[.010, .014]$ for animals, fruits and vegetables, respectively (and likewise for the corresponding $\Delta\text{BIC}$, $[-40,31]$, $[151,265]$, and $[193,280]$). Crucially, however, all quadratic solutions converged on concave curvature with strictly positive second derivatives $g''(x)=2a>0$ with CIs $[1.77,1.85]$, $[2.36,2.24]$, and $[2.22,2.31]$ (see Table~\ref{tab:model-parmeters-quadratic}) consistent with the universal law hypothesis. Finally, for the flexible GAM model, we found that it was unable to meaningfully improve on the exponential model despite its flexibility, with $\Delta R^2$ CIs given by $[-.006,-.002]$, $[-.008,-.004]$, $[-.006,-.002]$, and $\Delta \text{BIC}$ CIs given by $[-123,84]$, $[19,98]$, $[30,110]$, for animals, fruits and vegetables, respectively. \\   
As for the generalization data, we observed a similar pattern, namely, the exponential model outpeformed the linear ($\Delta R^2$ CIs, $[.019,.025]$, $[.014,.024]$, $[.012,.024]$, and $\Delta\text{BIC}$ CIs, $[276,366]$, $[135,242]$, $[103,214]$, for animals, fruits and vegetables, respectively). Likewise, the quadratic model performed on par with the exponential model ($\Delta R^2$ CIs, $[.001,.001]$, $[.002,.004]$, $[.008,.012]$, and $\Delta\text{BIC}$ CIs, $[4,17]$, $[28,45]$, $[68,109]$), and was strictly concave on all domains ($g''(x)=2a>0$ with CIs $[0.82,0.93]$, $[0.90,1.05]$, and $[0.99,1.13]$). Finally, the GAM model did not improve on the exponential model despite the additional degrees of freedom $\Delta R^2$ CIs, $[.002,.000]$, $[-.004,.000]$, $[-.009,.003]$, and $\Delta\text{BIC}$ CIs, $[133,163]$, $[120,151]$, $[75,128]$). Viewed together, these results provide direct evidence for the universal law of generalization.  

\begin{table}[ht]
  \caption{Full list of model evaluation metrics on the similarity tasks and their 95\% confidence intervals based on split-half bootstrap over trials with 100 repetitions.} \label{tab:model-performance-similarity}
\begin{tabular}{@{}lccccccccc@{}} 
\toprule
Category   & Model       & $R^2$   & $\delta R^2$ & $r_{md}$   & $\delta r_{md}$ & $r_c$    & $\delta r_c$ & $\Delta\text{BIC}$     & $\delta\Delta\text{BIC}$ \\
\midrule
Animals    & Exponential & 0.854 & 0.005   & 0.906 & 0.002   & 0.982 & 0.01   & 0.0      & 0.0      \\
Animals    & Linear      & 0.784 & 0.006   & 0.868 & 0.003   & 0.968 & 0.033  & 2782.613 & 138.579  \\
Animals    & Quadratic   & 0.854 & 0.005   & 0.906 & 0.002   & 0.98  & 0.008  & -4.387   & 35.575   \\
Animals    & GAM         & 0.858 & 0.005   & 0.907 & 0.003   & 0.983 & 0.008  & -19.403  & 103.188  \\
Fruits     & Exponential & 0.575 & 0.014   & 0.69  & 0.009   & 0.998 & 0.022  & 0.0      & 0.0      \\
Fruits     & Linear      & 0.441 & 0.014   & 0.615 & 0.008   & 0.933 & 0.048  & 1970.43  & 139.787  \\
Fruits     & Quadratic   & 0.563 & 0.015   & 0.683 & 0.009   & 0.991 & 0.021  & 207.94   & 56.923   \\
Fruits     & GAM         & 0.581 & 0.014   & 0.692 & 0.009   & 0.997 & 0.019  & 58.402   & 39.297   \\
Vegetables & Exponential & 0.645 & 0.01    & 0.75  & 0.007   & 0.99  & 0.009  & 0.0      & 0.0      \\
Vegetables & Linear      & 0.509 & 0.01    & 0.679 & 0.005   & 0.902 & 0.015  & 2321.842 & 136.625  \\
Vegetables & Quadratic   & 0.633 & 0.011   & 0.743 & 0.007   & 0.983 & 0.009  & 236.349  & 43.698   \\
Vegetables & GAM         & 0.649 & 0.01    & 0.752 & 0.007   & 0.992 & 0.008  & 69.554   & 39.957   \\
\bottomrule
\end{tabular}
\begin{tablenotes}
\small
\item Note:  The measures are: $R^2$ coefficient of determination, $r_{md}$ model-data Pearson correlation, $r_c$ model-data correlation corrected for attenuation, and $\Delta\text{BIC}$ the Bayesian Information Criterion (BIC) relative to the exponential model in each category. $\delta$ indicates 95\% confidence error (i.e., $\delta X = 1.96 \cdot \sigma_X$ where $\sigma_X$ is the standard deviation of $X$). See Methods for full details.
\end{tablenotes}
\end{table}

\begin{table}
  \caption{Full list of model evaluation metrics on the generalization tasks and their 95\% confidence intervals based on split-half bootstrap over trials with 100 repetitions.} \label{tab:model-performance-generalization}
\begin{tabular}{@{}lcccccccccc@{}} 
\toprule
Category   & Model       & $R^2$   & $\delta R^2$ & $r_{md}$   & $\delta r_{md}$ & $r_c$    & $\delta r_c$ & $\Delta\text{BIC}$     & $\delta\Delta\text{BIC}$ \\
\midrule
Animals    & Exponential & 0.522 & 0.031   & 0.655 & 0.01    & 0.951 & 0.015  & 0.0      & 0.0      \\
Animals    & Linear      & 0.5   & 0.032   & 0.641 & 0.01    & 0.946 & 0.026  & 320.879  & 44.845   \\
Animals    & Quadratic   & 0.521 & 0.031   & 0.656 & 0.01    & 0.95  & 0.014  & 10.57    & 6.792    \\
Animals    & GAM         & 0.523 & 0.031   & 0.656 & 0.01    & 0.951 & 0.015  & 148.101  & 14.716   \\
Fruits     & Exponential & 0.32  & 0.023   & 0.459 & 0.012   & 0.875 & 0.023  & 0.0      & 0.0      \\
Fruits     & Linear      & 0.301 & 0.022   & 0.449 & 0.012   & 0.863 & 0.028  & 188.706  & 53.238   \\
Fruits     & Quadratic   & 0.316 & 0.023   & 0.458 & 0.012   & 0.872 & 0.023  & 36.861   & 8.414    \\
Fruits     & GAM         & 0.322 & 0.023   & 0.46  & 0.012   & 0.879 & 0.024  & 135.5    & 15.998   \\
Vegetables & Exponential & 0.202 & 0.016   & 0.265 & 0.016   & 0.928 & 0.047  & 0.0      & 0.0      \\
Vegetables & Linear      & 0.184 & 0.014   & 0.263 & 0.016   & 0.903 & 0.048  & 158.541  & 55.12    \\
Vegetables & Quadratic   & 0.192 & 0.015   & 0.261 & 0.016   & 0.918 & 0.048  & 88.4     & 20.665   \\
Vegetables & GAM         & 0.209 & 0.016   & 0.27  & 0.016   & 0.933 & 0.045  & 101.458  & 26.378  \\
\bottomrule
\end{tabular}
\begin{tablenotes}
\small
\item Note:  See Table~\ref{tab:model-performance-similarity} for definitions of the various evaluation metrics.
\end{tablenotes}
\end{table}

\section{Discussion}
Shepard's universal law of generalization stands out as a theoretical claim about cognition in its intended scope, covering all intelligent entities and all stimuli. However, previous work had only evaluated it using relatively small, simple sets of stimuli. We assessed its performance in large sets of high-dimensional natural images comprising more than 600,000 human judgments. Our results provide robust evidence for the validity of the universal law and extend its long research tradition into rich naturalistic domains. By analyzing both similarity and generalization judgments, we also confirmed that generalization over blank properties and default similarity judgments indeed capture shared sources of variance even when the dimensionality of the space is particularly large. 

There are a number of limitations of the present work that can be further addressed by future research. First, our population was limited to online US participants to allow for efficient scaling of data crowdsourcing. However, cross-cultural research is necessary in order to evaluate the extent to which our findings generalize beyond US populations and English speakers \parencite{blasi2022over}. Nevertheless, the fact that we focused specifically on widespread natural categories should facilitate such an investigation. Second, in the present work we restricted ourselves to the visual modality, but one could equally consider natural categories in other primary modalities like the auditory and audio-visual (e.g., environmental sounds and scenes). While perhaps not as common as images, large behavioral datasets over such domains are becoming increasingly more accessible due to the growing interest in multi-modal models in the machine learning community (see e.g., \cite{gemmeke2017audio}; \cite{marjieh2022words}). Third, future work could explore how the results of our generalization analysis vary when other blank properties are considered. Indeed, one might expect that different blank properties may activate different forms of inductive reasoning \parencite{kemp2009structured} as well as inter-subject variation. The extent to which these too support the universal law of generalization is an open question that requires further investigation. Finally, naturalistic stimuli provide much more space than artificial stimuli for interrogating the relationship between generalization and similarity. Our results showed a significant correlation between similarity and generalization, but it varied significantly across domains. This raises questions such as what features of a complex stimulus people rely on when generalizing from one stimulus to another, and how their weights differ when people evaluate similarity. We hope to engage with these questions in future work.   

More broadly, our work showcases the prospects of scaling up psychological research, providing unprecedented precision for tests of foundational hypotheses in cognitive science, as well as new avenues for exploration of naturalistic stimuli. If our goal is to identify universal psychological principles underlying human cognition, being able to test those principles in naturalistic settings is essential to making strong claims about their universality. Finding that the universal law of generalization holds for natural images provides support for its use as a component of other cognitive models applied to these rich and complex stimuli (e.g., \cite{battleday2020capturing}; \cite{sanders2020training}), laying the groundwork for more extensive deployment and testing of models of human behavior based on psychological theory. \\

\noindent\textbf{Competing Interests.}
The authors declare no competing interests.
    
\noindent\textbf{Acknowledgments.}
This work was supported by grant 61454 from the John Templeton Foundation. 

\printbibliography

\appendix

\section{Supplementary Information}

\begin{figure}[htp]
  \centering
  \includegraphics[width=\linewidth]{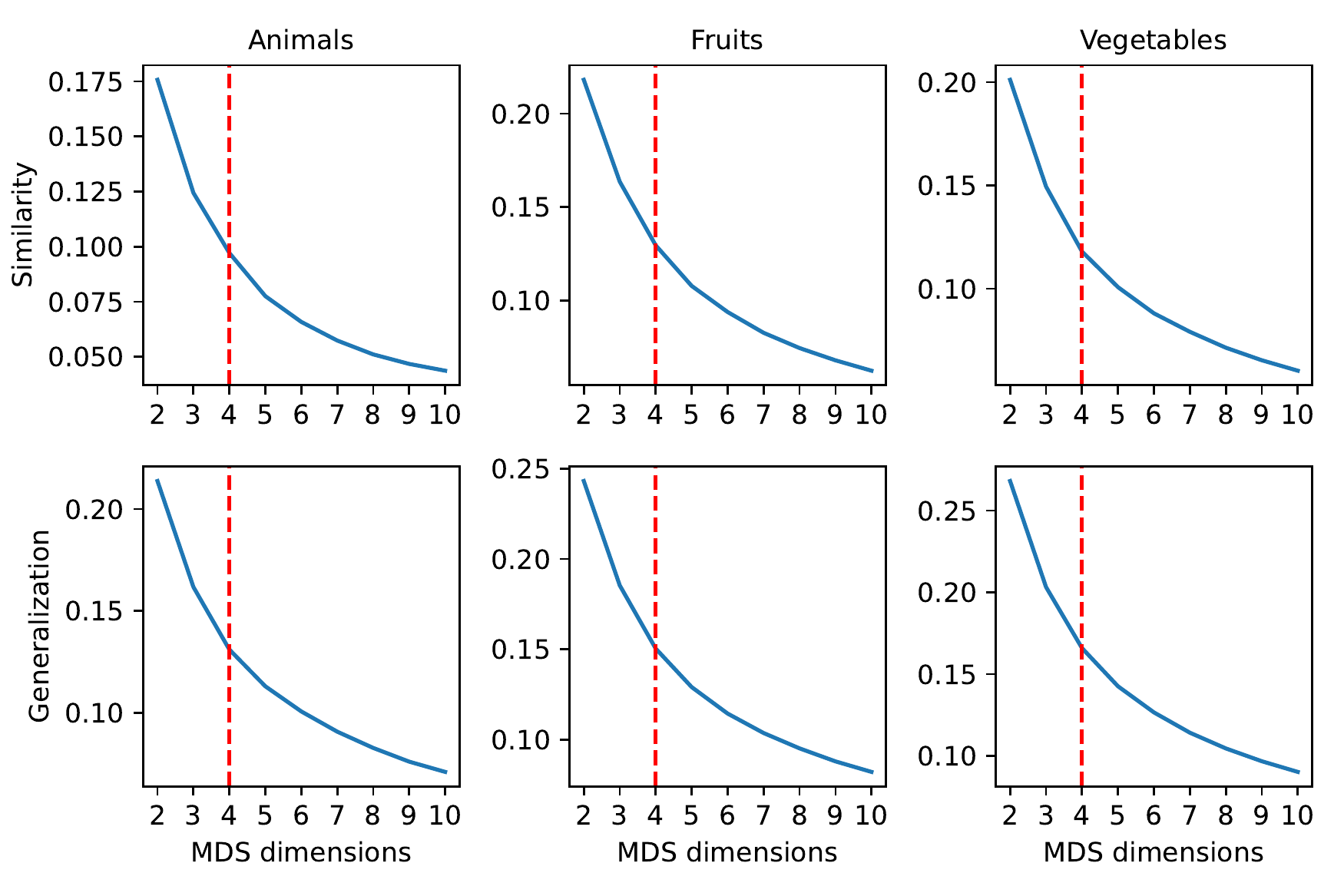}
%   \vspace{-4mm}
  \caption{Non-metric MDS stress curves for the various domains considered in the present work as a function of MDS dimensions. Based on this analysis we selected the first dimensionality ($d=4$, marked in dashed red vertical lines) for which all stress values across all datasets dropped below $0.2$ (a standard threshold above which MDS fit is deemed poor \parencite{kruskal1964multidimensional}).}
  \label{fig:stress-curves}
\end{figure}

\begin{figure}[ht]
  \centering
  \includegraphics[width=\linewidth]{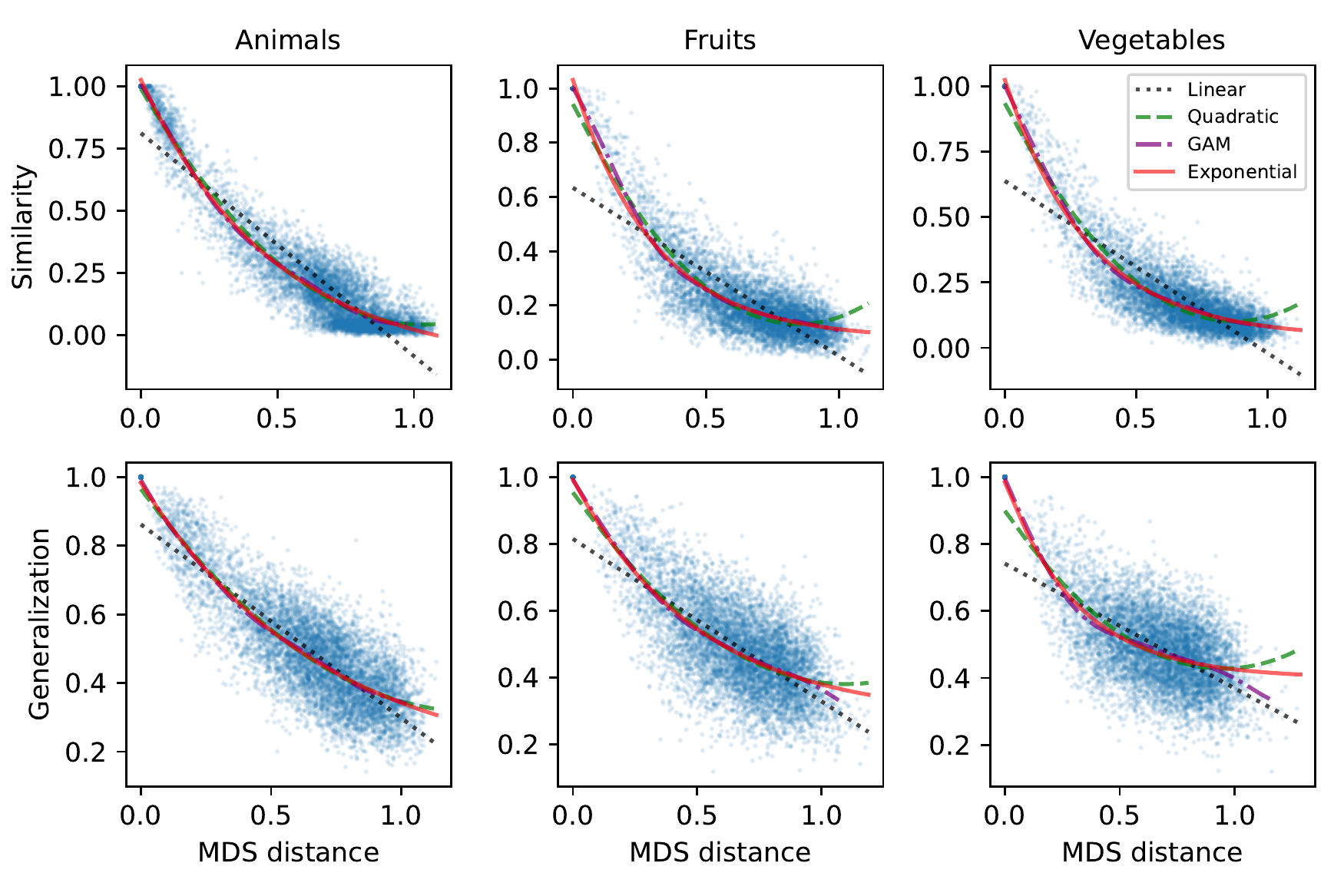}
%   \vspace{-4mm}
  \caption{Raw generalization gradients across domains of natural images and tasks with the optimal fitted models overlaid.}
  \label{fig:universal-law-raw}
\end{figure}

\begin{table}[ht]
\caption{Exponential model parameters of the form $g(x)=a e^{-bx} + c$. Errors ($\delta$) indicate 95\% confidence intervals bootstrapped over trials with 100 repetitions.}
\label{tab:model-parmeters}
\begin{tabular}{@{}lccccccc@{}}        
\toprule
Task           & Category   & $a$     & $\delta a$ & $b$     & $\delta b$ & $c$      & $\delta c$ \\
\midrule
Similarity     & Animals    & 1.179 & 0.009 & 2.022 & 0.043 & -0.138 & 0.009 \\
Similarity     & Fruits     & 0.969 & 0.008 & 3.243 & 0.075 & 0.072  & 0.005 \\
Similarity     & Vegetables & 1.005 & 0.007 & 3.08  & 0.056 & 0.032  & 0.005 \\
Generalization & Animals    & 0.844 & 0.018 & 1.525 & 0.085 & 0.154  & 0.022 \\
Generalization & Fruits     & 0.711 & 0.012 & 2.029 & 0.114 & 0.28   & 0.016 \\
Generalization & Vegetables & 0.604 & 0.009 & 2.916 & 0.144 & 0.378  & 0.01 \\ 
\bottomrule
\end{tabular}
\end{table}

\begin{table}[ht]
\caption{Linear model parameters of the form $g(x)=a x + b$. Errors ($\delta$) indicate 95\% confidence intervals bootstrapped over trials with 100 repetitions.}
\label{tab:model-parmeters-linear}
\begin{tabular}{@{}lccccc@{}}        
\toprule
Task           & Category       & $a$     & $\delta a$ & $b$     & $\delta b$ \\
\midrule
Similarity     & Animals    & -0.917 & 0.005 & 0.827 & 0.003 \\
Similarity     & Fruits     & -0.629 & 0.008 & 0.639 & 0.006 \\
Similarity     & Vegetables & -0.671 & 0.007 & 0.646 & 0.005 \\
Generalization & Animals    & -0.568 & 0.007 & 0.86  & 0.005 \\
Generalization & Fruits     & -0.484 & 0.008 & 0.807 & 0.006 \\
Generalization & Vegetables & -0.39  & 0.009 & 0.745 & 0.006 \\ 
\bottomrule
\end{tabular}
\end{table}

\begin{table}[ht]
\caption{Quadratic model parameters of the form $g(x)=a x^2 + b x + c$. Errors ($\delta$) indicate 95\% confidence intervals bootstrapped over trials with 100 repetitions.}
\label{tab:model-parmeters-quadratic}
\begin{tabular}{@{}lccccccc@{}}        
\toprule
Task           & Category   & $a$     & $\delta a$ & $b$     & $\delta b$ & $c$      & $\delta c$ \\
\midrule
Similarity     & Animals    & 0.906 & 0.021 & -1.872 & 0.023 & 1.007 & 0.005 \\
Similarity     & Fruits     & 1.152 & 0.03  & -1.951 & 0.033 & 0.957 & 0.009 \\
Similarity     & Vegetables & 1.134 & 0.024 & -1.968 & 0.026 & 0.956 & 0.007 \\
Generalization & Animals    & 0.436 & 0.027 & -1.069 & 0.03  & 0.977 & 0.007 \\
Generalization & Fruits     & 0.488 & 0.036 & -1.058 & 0.04  & 0.951 & 0.01  \\
Generalization & Vegetables & 0.528 & 0.035 & -1.017 & 0.039 & 0.906 & 0.01 \\
\bottomrule
\end{tabular}
\end{table}

\begin{table}[ht]
  \caption{Complementary list of evaluation measures on the similarity tasks and their 95\% confidence intervals based on split-half bootstrap over trials with 100 repetitions.} \label{tab:model-splithalf-similarity}
\begin{tabular}{@{}lccccccc@{}} 
\toprule
Category   & Model       & $r_{dd}$   & $\delta r_{dd}$ & $r_{mm}$   & $\delta r_{mm}$ & $r_{md}$   & $\delta r_{md}$ \\
\midrule
Animals    & Exponential & 0.89  & 0.004   & 0.956 & 0.019   & 0.906 & 0.002   \\
Animals    & Linear      & 0.89  & 0.004   & 0.905 & 0.063   & 0.868 & 0.003   \\
Animals    & Quadratic   & 0.89  & 0.004   & 0.96  & 0.016   & 0.906 & 0.002   \\
Animals    & GAM         & 0.89  & 0.004   & 0.958 & 0.016   & 0.907 & 0.003   \\
Fruits     & Exponential & 0.591 & 0.011   & 0.808 & 0.041   & 0.69  & 0.009   \\
Fruits     & Linear      & 0.591 & 0.011   & 0.737 & 0.083   & 0.615 & 0.008   \\
Fruits     & Quadratic   & 0.591 & 0.011   & 0.803 & 0.039   & 0.683 & 0.009   \\
Fruits     & GAM         & 0.591 & 0.011   & 0.815 & 0.036   & 0.692 & 0.009   \\
Vegetables & Exponential & 0.64  & 0.01    & 0.897 & 0.018   & 0.75  & 0.007   \\
Vegetables & Linear      & 0.64  & 0.01    & 0.884 & 0.033   & 0.679 & 0.005   \\
Vegetables & Quadratic   & 0.64  & 0.01    & 0.893 & 0.019   & 0.743 & 0.007   \\
Vegetables & GAM         & 0.64  & 0.01    & 0.898 & 0.017   & 0.752 & 0.007   \\
\bottomrule
\end{tabular}
\begin{tablenotes}
Note:  The measures are: $r_{dd}$ data-data Pearson correlation, $r_{mm}$ model-model Pearson correlation, and $r_{md}$ model-data Pearson correlation.  $\delta$ indicates 95\% confidence error (i.e., $\delta X = 1.96 \cdot \sigma_X$ where $\sigma_X$ is the standard deviation of $X$).
\end{tablenotes}
\end{table}

\begin{table}
  \caption{Complementary list of evaluation measures on the generalization tasks and their 95\% confidence intervals based on split-half bootstrap over trials with 100 repetitions.} \label{tab:model-splithalf-generalization}
\begin{tabular}{@{}lccccccc@{}} 
\toprule
Category   & Model       & $r_{dd}$   & $\delta r_{dd}$ & $r_{mm}$   & $\delta r_{mm}$ & $r_{md}$   & $\delta r_{md}$ \\
\midrule
Animals    & Exponential & 0.577 & 0.011   & 0.823 & 0.037   & 0.655 & 0.01    \\
Animals    & Linear      & 0.577 & 0.011   & 0.796 & 0.056   & 0.641 & 0.01    \\
Animals    & Quadratic   & 0.577 & 0.011   & 0.825 & 0.036   & 0.656 & 0.01    \\
Animals    & GAM         & 0.577 & 0.011   & 0.824 & 0.037   & 0.656 & 0.01    \\
Fruits     & Exponential & 0.417 & 0.013   & 0.661 & 0.041   & 0.459 & 0.012   \\
Fruits     & Linear      & 0.417 & 0.013   & 0.649 & 0.049   & 0.449 & 0.012   \\
Fruits     & Quadratic   & 0.417 & 0.013   & 0.66  & 0.04    & 0.458 & 0.012   \\
Fruits     & GAM         & 0.417 & 0.013   & 0.658 & 0.042   & 0.46  & 0.012   \\
Vegetables & Exponential & 0.188 & 0.016   & 0.437 & 0.053   & 0.265 & 0.016   \\
Vegetables & Linear      & 0.188 & 0.016   & 0.454 & 0.061   & 0.263 & 0.016   \\
Vegetables & Quadratic   & 0.188 & 0.016   & 0.432 & 0.055   & 0.261 & 0.016   \\
Vegetables & GAM         & 0.188 & 0.016   & 0.447 & 0.054   & 0.27  & 0.016 \\
\bottomrule
\end{tabular}
\begin{tablenotes}
    Note:  See Table~\ref{tab:model-splithalf-similarity} for definition of the various metrics.
\end{tablenotes}
\end{table}

\section{Binned Analysis}
\begin{table}
\caption{Exponential model parameters of the form $g(x)=a e^{-bx} + c$. Errors ($\delta$) indicate 95\% confidence intervals bootstrapped over trials with 100 repetitions and $n=100$ bins.}
\label{tab:model-parmeters-binned}
\begin{tabular}{@{}lccccccc@{}}        
\toprule
Task           & Category   & $a$     & $\delta a$ & $b$     & $\delta b$ & $c$      & $\delta c$ \\
\midrule
Similarity     & Animals    & 1.154 & 0.017 & 2.194 & 0.101 & -0.097 & 0.021 \\
Similarity     & Fruits     & 1.059 & 0.028 & 3.396 & 0.14  & 0.071  & 0.01  \\
Similarity     & Vegetables & 1.069 & 0.022 & 3.195 & 0.129 & 0.032  & 0.009 \\
Generalization & Animals    & 0.844 & 0.044 & 1.525 & 0.201 & 0.153  & 0.056 \\
Generalization & Fruits     & 0.75  & 0.044 & 1.769 & 0.347 & 0.229  & 0.068 \\
Generalization & Vegetables & 0.608 & 0.036 & 2.546 & 0.661 & 0.344  & 0.063 \\
\bottomrule
\end{tabular}
\end{table}

\begin{table}
\caption{Linear model parameters of the form $g(x)=a x + b$. Errors ($\delta$) indicate 95\% confidence intervals bootstrapped over trials with 100 repetitions and $n=100$ bins.}
\label{tab:model-parmeters-binned-linear}
\begin{tabular}{@{}lccccc@{}}        
\toprule
Task           & Category       & $a$     & $\delta a$ & $b$     & $\delta b$ \\
\midrule
Similarity     & Animals    & -0.882 & 0.046 & 0.817 & 0.018 \\
Similarity     & Fruits     & -0.723 & 0.03  & 0.721 & 0.016 \\
Similarity     & Vegetables & -0.77  & 0.032 & 0.731 & 0.015 \\
Generalization & Animals    & -0.57  & 0.02  & 0.869 & 0.01  \\
Generalization & Fruits     & -0.491 & 0.02  & 0.819 & 0.012 \\
Generalization & Vegetables & -0.385 & 0.03  & 0.747 & 0.016 \\
\bottomrule
\end{tabular}
\end{table}

\begin{table}
\caption{Quadratic model parameters of the form $g(x)=a x^2 + b x + c$. Errors ($\delta$) indicate 95\% confidence intervals bootstrapped over trials with 100 repetitions and $n=100$ bins.}
\label{tab:model-parmeters-quadratic-binned}
\begin{tabular}{@{}lccccccc@{}}        
\toprule
Task           & Category   & $a$     & $\delta a$ & $b$     & $\delta b$ & $c$      & $\delta c$ \\
\midrule
Similarity     & Animals    & 0.958 & 0.033 & -1.919 & 0.033 & 1.008 & 0.008 \\
Similarity     & Fruits     & 1.203 & 0.086 & -2.057 & 0.093 & 0.992 & 0.024 \\
Similarity     & Vegetables & 1.189 & 0.077 & -2.067 & 0.081 & 0.984 & 0.019 \\
Generalization & Animals    & 0.43  & 0.072 & -1.063 & 0.073 & 0.974 & 0.015 \\
Generalization & Fruits     & 0.386 & 0.099 & -0.96  & 0.11  & 0.93  & 0.026 \\
Generalization & Vegetables & 0.37  & 0.172 & -0.841 & 0.189 & 0.858 & 0.043 \\
\bottomrule
\end{tabular}
\end{table}

\begin{table}
  \caption{Full list of model evaluation metrics on the similarity tasks and their 95\% confidence intervals based on split-half bootstrap over trials with 100 repetitions and $n=100$ bins.} \label{tab:model-performance-similarity-binned}
\begin{tabular}{@{}lccccccccc@{}} 
\toprule
Category   & Model       & $R^2$   & $\delta R^2$ & $r_{md}$   & $\delta r_{md}$ & $r_c$    & $\delta r_c$ & $\Delta\text{BIC}$     & $\delta\Delta\text{BIC}$ \\
\midrule
Animals    & Exponential & 0.994 & 0.002   & 0.997 & 0.001   & 0.999 & 0.001  & 0.0     & 0.0      \\
Animals    & Linear      & 0.911 & 0.014   & 0.954 & 0.005   & 0.956 & 0.005  & 257.735 & 27.564   \\
Animals    & Quadratic   & 0.993 & 0.002   & 0.996 & 0.001   & 0.998 & 0.001  & 16.567  & 20.269   \\
Animals    & GAM         & 0.997 & 0.001   & 0.998 & 0.001   & 1.0   & 0.0    & 21.103  & 40.222   \\
Fruits     & Exponential & 0.986 & 0.005   & 0.993 & 0.002   & 0.998 & 0.002  & 0.0     & 0.0      \\
Fruits     & Linear      & 0.812 & 0.018   & 0.902 & 0.005   & 0.906 & 0.006  & 226.335 & 31.753   \\
Fruits     & Quadratic   & 0.969 & 0.008   & 0.984 & 0.003   & 0.989 & 0.003  & 71.777  & 25.337   \\
Fruits     & GAM         & 0.991 & 0.005   & 0.994 & 0.002   & 1.0   & 0.001  & 39.543  & 30.356   \\
Vegetables & Exponential & 0.991 & 0.004   & 0.995 & 0.001   & 0.999 & 0.001  & 0.0     & 0.0      \\
Vegetables & Linear      & 0.837 & 0.014   & 0.915 & 0.004   & 0.919 & 0.004  & 251.392 & 35.803   \\
Vegetables & Quadratic   & 0.979 & 0.005   & 0.989 & 0.002   & 0.993 & 0.002  & 72.835  & 31.745   \\
Vegetables & GAM         & 0.993 & 0.004   & 0.996 & 0.002   & 1.0   & 0.001  & 52.324  & 28.149   \\
\bottomrule
\end{tabular}
\begin{tablenotes}
\small
\item Note: The measures are: $R^2$ coefficient of determination, $r_{md}$ model-data Pearson correlation, $r_c$ model-data correlation corrected for attenuation, and $\Delta\text{BIC}$ the Bayesian Information Criterion (BIC) relative to the exponential model in each category. $\delta$ indicates 95\% confidence error (i.e., $\delta X = 1.96 \cdot \sigma_X$ where $\sigma_X$ is the standard deviation of $X$). See Methods for full details.
\end{tablenotes}
\end{table}

\begin{table}
  \caption{Full list of model evaluation metrics on the generalization tasks and their 95\% confidence intervals based on split-half bootstrap over trials with 100 repetitions and $n=100$ bins.} \label{tab:model-performance-generalization-binned}
\begin{tabular}{@{}lcccccccccc@{}} 
\toprule
Category   & Model       & $R^2$   & $\delta R^2$ & $r_{md}$   & $\delta r_{md}$ & $r_c$    & $\delta r_c$ & $\Delta\text{BIC}$     & $\delta\Delta\text{BIC}$ \\
\midrule
Animals    & Exponential & 0.989 & 0.01    & 0.994 & 0.004   & 1.0   & 0.003  & 0.0     & 0.0      \\
Animals    & Linear      & 0.951 & 0.02    & 0.975 & 0.007   & 0.98  & 0.007  & 125.869 & 39.802   \\
Animals    & Quadratic   & 0.989 & 0.008   & 0.994 & 0.003   & 0.999 & 0.003  & 3.683   & 16.468   \\
Animals    & GAM         & 0.992 & 0.004   & 0.993 & 0.006   & 1.0   & 0.001  & 53.514  & 53.277   \\
Fruits     & Exponential & 0.97  & 0.016   & 0.985 & 0.006   & 0.997 & 0.006  & 0.0     & 0.0      \\
Fruits     & Linear      & 0.932 & 0.02    & 0.966 & 0.006   & 0.978 & 0.007  & 66.156  & 44.537   \\
Fruits     & Quadratic   & 0.965 & 0.017   & 0.982 & 0.007   & 0.994 & 0.006  & 13.808  & 9.359    \\
Fruits     & GAM         & 0.982 & 0.01    & 0.986 & 0.009   & 1.0   & 0.003  & 40.117  & 46.13    \\
Vegetables & Exponential & 0.935 & 0.052   & 0.965 & 0.021   & 0.991 & 0.017  & 0.0     & 0.0      \\
Vegetables & Linear      & 0.888 & 0.051   & 0.945 & 0.018   & 0.97  & 0.02   & 39.1    & 54.258   \\
Vegetables & Quadratic   & 0.924 & 0.044   & 0.958 & 0.023   & 0.985 & 0.016  & 12.989  & 18.99    \\
Vegetables & GAM         & 0.974 & 0.015   & 0.964 & 0.039   & 1.0   & 0.005  & 9.844   & 63.035 \\
\bottomrule
\end{tabular}
\begin{tablenotes}
\small
\item Note:  See Table~\ref{tab:model-performance-similarity} for definitions of the various evaluation metrics.
\end{tablenotes}
\end{table}

\begin{table}
  \caption{Complementary list of evaluation measures on the similarity tasks and their 95\% confidence intervals based on split-half bootstrap over trials with 100 repetitions and $n=100$ bins.} \label{tab:model-splithalf-similarity-binned}
\begin{tabular}{@{}lccccccc@{}} 
\toprule
Category   & Model       & $r_{dd}$   & $\delta r_{dd}$ & $r_{mm}$   & $\delta r_{mm}$ & $r_{md}$   & $\delta r_{md}$ \\
\midrule
Animals    & Exponential & 0.996 & 0.002   & 1.0   & 0.0     & 0.997 & 0.001   \\
Animals    & Linear      & 0.996 & 0.002   & 1.0   & 0.0     & 0.954 & 0.005   \\
Animals    & Quadratic   & 0.996 & 0.002   & 1.0   & 0.0     & 0.996 & 0.001   \\
Animals    & GAM         & 0.996 & 0.002   & 0.999 & 0.001   & 0.998 & 0.001   \\
Fruits     & Exponential & 0.99  & 0.005   & 1.0   & 0.0     & 0.993 & 0.002   \\
Fruits     & Linear      & 0.99  & 0.005   & 1.0   & 0.0     & 0.902 & 0.005   \\
Fruits     & Quadratic   & 0.99  & 0.005   & 1.0   & 0.0     & 0.984 & 0.003   \\
Fruits     & GAM         & 0.99  & 0.005   & 0.999 & 0.001   & 0.994 & 0.002   \\
Vegetables & Exponential & 0.992 & 0.004   & 1.0   & 0.0     & 0.995 & 0.001   \\
Vegetables & Linear      & 0.992 & 0.004   & 1.0   & 0.0     & 0.915 & 0.004   \\
Vegetables & Quadratic   & 0.992 & 0.004   & 1.0   & 0.0     & 0.989 & 0.002   \\
Vegetables & GAM         & 0.992 & 0.004   & 0.999 & 0.001   & 0.996 & 0.002 \\
\bottomrule
\end{tabular}
\begin{tablenotes}
Note:  The measures are: $r_{dd}$ data-data Pearson correlation, $r_{mm}$ model-model Pearson correlation, and $r_{md}$ model-data Pearson correlation.  $\delta$ indicates 95\% confidence error (i.e., $\delta X = 1.96 \cdot \sigma_X$ where $\sigma_X$ is the standard deviation of $X$).
\end{tablenotes}
\end{table}

\begin{table}
  \caption{Complementary list of evaluation measures on the generalization tasks and their 95\% confidence intervals based on split-half bootstrap over trials with 100 repetitions and $n=100$ bins.} \label{tab:model-splithalf-generalization-binned}
\begin{tabular}{@{}lccccccc@{}} 
\toprule
Category   & Model       & $r_{dd}$   & $\delta r_{dd}$ & $r_{mm}$   & $\delta r_{mm}$ & $r_{md}$   & $\delta r_{md}$ \\
\midrule
Animals    & Exponential & 0.989 & 0.007   & 1.0   & 0.001   & 0.994 & 0.004   \\
Animals    & Linear      & 0.989 & 0.007   & 1.0   & 0.0     & 0.975 & 0.007   \\
Animals    & Quadratic   & 0.989 & 0.007   & 1.0   & 0.001   & 0.994 & 0.003   \\
Animals    & GAM         & 0.989 & 0.007   & 0.998 & 0.006   & 0.993 & 0.006   \\
Fruits     & Exponential & 0.976 & 0.014   & 1.0   & 0.001   & 0.985 & 0.006   \\
Fruits     & Linear      & 0.976 & 0.014   & 1.0   & 0.0     & 0.966 & 0.006   \\
Fruits     & Quadratic   & 0.976 & 0.014   & 0.999 & 0.002   & 0.982 & 0.007   \\
Fruits     & GAM         & 0.976 & 0.014   & 0.996 & 0.008   & 0.986 & 0.009   \\
Vegetables & Exponential & 0.95  & 0.042   & 0.998 & 0.004   & 0.965 & 0.021   \\
Vegetables & Linear      & 0.95  & 0.042   & 1.0   & 0.0     & 0.945 & 0.018   \\
Vegetables & Quadratic   & 0.95  & 0.042   & 0.995 & 0.011   & 0.958 & 0.023   \\
Vegetables & GAM         & 0.95  & 0.042   & 0.979 & 0.037   & 0.964 & 0.039 \\
\bottomrule
\end{tabular}
\begin{tablenotes}
    Note:  See Table~\ref{tab:model-splithalf-similarity-binned} for definition of the various metrics.
\end{tablenotes}
\end{table}
% As shown in Figure~\ref{fig:Figure2}, these results are impressive. \lipsum[20]

% \begin{figure}
%     \caption{This is my second figure caption.}
%     \includegraphics[bb=0in 0in 2.5in 2.5in, height=2.5in, width=2.5in]{Figure1.pdf}
%     \label{fig:Figure2}
% \end{figure}

% \lipsum[21]
% \section{Pilot Data}
% \label{app:surveydata}

% The detailed results are shown in Table~\ref{tab:DeckedTable}. \lipsum[22]

% \begin{table}
%   \begin{threeparttable}
%     \caption{A More Complex Decked Table}
%     \label{tab:DeckedTable}
%     \begin{tabular}{@{}lrrr@{}}         \toprule
%     Distribution type  & \multicolumn{2}{l}{Percentage of} & Total number   \\
%                        & \multicolumn{2}{l}{targets with}  & of trials per  \\
%                        & \multicolumn{2}{l}{segment in}    & participant    \\ \cmidrule(r){2-3}
%                                     &  Onset  &  Coda            &          \\ \midrule
%     Categorical -- onset\tabfnm{a}  &    100  &     0            &  196     \\
%     Probabilistic                   &     80  &    20\tabfnm{*}  &  200     \\
%     Categorical -- coda\tabfnm{b}   &      0  &   100\tabfnm{*}  &  196     \\ \midrule
%     \end{tabular}
%     \begin{tablenotes}[para,flushleft]
%         {\small
%             \textit{Note.} All data are approximate.

%             \tabfnt{a}Categorical may be onset.
%             \tabfnt{b}Categorical may also be coda.

%             \tabfnt{*}\textit{p} < .05.
%             \tabfnt{**}\textit{p} < .01.
%          }
%     \end{tablenotes}
%   \end{threeparttable}
% \end{table}

% \lipsum[23]

% \printbibliography
\end{document}